# It is not all automated and false: An analysis of suppressing the #MahsaAmini movement on Persian Twitter

**Hossein Kermani**
Political Communication Group,
Department of Communication,
University of Vienna
Hossein.kermani@univie.ac.at

**Fatemeh Oudlajani**
Allameh Tabataba'i University
fatemehoudlajani@gmail.com

**Pardis Yarahmadi**
Queensland University of Technology
P.Yarahmadi@hdr.qut.edu.au

**Hamideh Mahdi Soltani**
Allameh Tabataba'i University
ms.soltani1991@gmail.com

**Mohammad Makki**
University of Wollongong, Australia
mmakki@uow.edu.au

**Zahra HosseiniKhoo**
University of Vienna
z.hosseinikhoo@gmail.com

### Abstract

This paper investigates the actors and practices of suppressing the #MahsaAmini movement on Persian Twitter. Despite the growing body of studies into social media suppression, the literature remains relatively Western-centered. In addition, extant research is mainly restricted to false information and social bots as the primary practice and agents of suppression. This paper pushes forward this line of inquiry by focusing on an understudied context and following an exploratory and discursive approach to social media suppression. We identified and analyzed pro-regime users who shared the most popular tweets during the movement (N=1,388) and extracted their tweets (N=6,170). Findings reveal that suppressive agents during the movement were mainly elite/ordinary, not-coordinated, and real individuals. The frequency of suspicious, fake, and automated accounts was minor, contradicting the existing literature. In addition, results indicate that incivility was the main suppressive tactic. Implications for further studies and contributions to the existing literature will also be further discussed.

## 1. Introduction

In September 2022, a substantial anti-regime movement, known as #MahsaAmini or Women, Life, Liberty (WLL), spurred the streets and social media in Iran in less than a week. In addition to unprecedented street demonstrations, many people went on Twitter to show their frustration with the regime and raise their voices against its policies. #MahsaAmini broke a record in Twitter history with more than 500 million tweets from September 15th to December 15th, 2022. As the movement gained remarkable international attention, the Iranian regime invested heavily to suppress it. Scholars have recently studied suppressing social media movements and other counter-political activism in different countries. The 2016 U.S. presidential election, as well as the BREXIT, put the ways through which non-democratic forces manipulate and repress online social movements at the center of attention (Bastos & Mercea, 2019; Howard et al., 2018). However, the extant research is relatively Western-centered, focusing mainly on the U.S. (Allcott & Gentzkow, 2017; Benkler et al., 2018; Howard et al., 2018; Linvill & Warren, 2020; Lukito, 2020) and some European democracies like England (Bastos & Mercea, 2019; Howard & Kollanyi, 2016) and Germany (Muller & Schwarz, 2017; Rogers & Niederer, 2020). While there are some works focused on non-democratic societies (Forelle et al., 2015; Maweu, 2019; Owen Jones, 2022), we need more research on such understudied contexts to understand better the breadth, scale, and dynamics of disinformation in contemporary societies. This study contributes to this growing body of literature by providing fresh insights from a non-Western, understudied context: Iran.

In addition, the existing literature mainly concentrates on false information and social bots as practices and agents of social media suppression (Lukito, 2020; Woolley & Howard, 2018). Alongside Starbird (2019), we argue that a broader understanding of this phenomenon is required to understand how social media movements are suppressed in different contexts. In this sense, we follow an exploratory approach to investigate potential bad actors, i.e., agents who try to infiltrate organic social media communication, and their practices discursively.

In order to address these gaps, our empirical analyses have focused on a dataset of 78,088 Persian tweets shared during the first months of the movement. We follow KhosraviNik's (2017) approach to Social Media Critical Discourse Studies (SM-CDS) to qualitatively and discursively analyze the research sample. We identified 1,388 pro-regime users and extracted all their tweets to investigate how they tried to exterminate the anti-regime movement. We start by reviewing the existing literature in the next section. Then, we will discuss social media and politics in Iran. Next, we will delineate our methodological approach and present the research findings. Finally, we will discuss the contribution of this research to the growing body of literature on social media suppression.

## 2. Computational propaganda and social media suppression

The extant literature on social media suppression (SMS) revolves around the concept of computational propaganda (CP), defined as "the assemblage of social media platforms, autonomous agents, and big data tasked with the manipulation of public opinion" (Woolley & Howard, 2016, p. 4886). From another perspective, computational propaganda could be understood as the use of computational means to automate propaganda tactics in deliberately subverting symbols, appealing to our baser emotions and prejudices, and bypassing rational thought to achieve the specific goals of its promoters (Howard et al., 2018). CP is a broad concept that can be studied only by investigating its subsets. We argue that CP consists of two main parts: actors and practices (i.e., tactics). Actors, i.e., the agents of CP, are responsible for orchestrating malicious campaigns in front and back stages. Scholars like Keller et al. (2020) differentiate between actors and principles. Principles are those who order and manage the campaigns, and actors are the front-end soldiers who perform the orders. Practices are underlying ways to manipulate narratives and change the facts.

While scholars have discussed a variety of actors and practices (Owen Jones, 2022; Rogers & Niederer, 2020), the extant research empirically focuses on sharing false information as the primary practice and social bots as the main actors of CP to a great extent. False information is a term that refers to all types of problematic messages and information that attempt to deceive and manipulate public opinion. Disinformation and misinformation are its main parts. They are both false, but the latter is unintentional, while the former is created for intentional disruption and harm (Rogers & Niederer, 2020). While other terms like fake news have become a buzzword in this field, researchers are cautious about using them as they lack theoretical sensitivity and are overloaded (Bennett & Livingston, 2018; Freelon & Wells, 2020).

The existing literature unanimously emphasizes that false information (FI) threatens democratic practices in different contexts, from democracies to non-democratic societies by sowing discord, amplifying the divisions that polarize society, making the flow of discussion and communication toxic, and manipulating facts (Chadwick & Stanyer, 2022; Diaz Ruiz & Nilsson, 2023; Ferreira, 2022; Hameleers, 2022; Rogers & Niederer, 2020). We add another practice that could be detrimental independently and in combination with false information: online incivility. While non-democratic forces have used online incivility to, e.g., silence dissident citizens, CP literature has not accounted for it significantly.

Coe et al. (2014) understand incivility as features of discussion that convey an unnecessarily disrespectful tone toward the discussion forum, its participants, or its topics (p. 660). They suggest a typology of uncivil speech: Name-calling, Aspersion, Vulgarity, Lying, and the pejorative of speech (p. 661). This typology helps us delineate the different levels of incivility, whilst other buzzwords like hate speech do not. The existing literature has studied incivility in personal communication (Blau & Andersson, 2005; Forni, 2002) and political discourse (Brooks & Geer, 2007; Fridkin & Kenney, 2008; Mutz, 2007). On social media, Shmargad et al. (2022) investigate how social norms affect the prevalence and presence of online incivility. In another study, Rowe (2015) examines the effects of anonymity on the uncivil behavior of those who post on news outlets pages on Facebook. In a study more relevant to the current research, Bratslavsky et al. (2019) discuss how incivility is a strategy related to power and domination, particularly from the former president of the United States, Donald Trump. They propose that Twitter is part of an infrastructure of incivility through which structures and discursive mechanisms contribute to a devaluation of normative democratic discourses. This is, in fact, the reason why many scholars attribute incivility to far-right and partisan politics (Van Spanje & De Vreese, 2014).

There is a discursive affinity between FI and incivility. Both are partisan and based on projection, prejudice, categorization, and dehumanization. Hameleers et al. (2022) discuss how the aims and consequences of spreading false information and incivility align. The targeted spread of FI and the expression of uncivil speech may involve (des)identification processes in which groups are attacked and ascribed negative traits (Hameleers et al., 2022, p. 1599). The extant research suggests that exposure to uncivil political discourse in the media erodes political trust, engenders more negative assessments of political institutions and actors, decreases the perceived legitimacy of political figures, and triggers increased emotional response (e.g., Forgette & Morris, 2006; Fridkin & Kenney, 2008; Mutz, 2007; Anderson et al., 2018). In addition, we postulate that suppressive tactics are not limited to the above-discussed practices. We will follow an exploratory approach to identify all possible tactics for suppressing the #MahsaAmini movement. In this sense, we will go beyond the extant research, which is remarkably limited to CP and FI.

Furthermore, most research has focused on social bots as harmful agents. Social bots are software programs or agents created to perform simple, repetitive, typically text-based tasks (Woolley & Howard, 2018). A primary task of social bots is undoubtedly spreading disinformation (Bessi & Ferrara, 2016; Cai et al., 2023), but we should keep two points in mind. First, being a bot does not necessarily mean they are always active in disinformation campaigns (Keller et al., 2020). Also, other forms of Inauthentic Social Media Accounts (ISMA) are active in disinformation operations. We prefer to focus on ISMA instead of social bots as actors of suppression. ISMA are users who do not genuinely and honestly engage in social media activities. They are malicious actors tasked with infiltrating and manipulating public discourse. There are several forms of ISMA: bots, trolls, and cyborgs. Trolls are humans hired to disseminate false information and engage with disinformation to increase the virality and prominence of dishonest information (e.g., Starbird, 2019). Cyborgs have the behavior patterns of both bots and humans and are partially human-operated (Chu et al., 2012). Furthermore, we hypothesize that even real identities, e.g., ordinary citizens or politicians, could be the agents of suppression on social media. This point is largely neglected in the existing literature.

We make another theoretical contribution to the existing literature. Since social bots are at the center of attention, extant research uses different bot detection methods to measure the probability of an account being a bot (Martini et al., 2021). As discussed above, restricting our focus to social bots could be misleading. As a result, we suggest using another measure, Inauthentic Coordinated Behavior probability (ICB). ICB shows the probability that a social media account is part of an inauthentic coordinated campaign. Thus, we go beyond the extant research again to explore all types of accounts involved in suppressing the #MahsaAmini movement.

## 3. Social media suppression beyond the West: the case of Iran

As discussed above, the extant research on SMS is Western-centered. The emphasis on Western democracies could cause theoretical and empirical shortcomings. As we discussed before, the existing literature concentrates on social bots as bad agents and false information as the main practice of SMS. In Western democracies, it is not common for real individuals like politicians or journalists to work as elite false information disseminators. They might, but in such cases, they will be heavily criticized by the public and media. In addition, there are legal settings to sue bad actors. Thus, many bad campaigners prefer to use social bots to propel their malicious aims in unknown ways. The concentration of the extant research on false information could be contingent on the fact that such misleading messages are more accessible to identify and detect computationally. However, these arguments remain hypothetical regarding the need for empirical studies from non-democratic societies. Studying Iran as an authoritarian regime could enhance our knowledge of the nuances and dynamics of SMS in diverse and particularly understudied contexts.

Iran is an authoritarian country using cyber-armies and different tactics to suppress anti-regime movements vastly (Kargar & Rauchfleisch, 2019; Kermani, 2023). While major social media like Twitter are blocked, Iranians continue to use them by employing VPNs and proxies. That is why the regime has developed various deception operations to corrupt social media protests from within. There is clear evidence of the regime's notorious social media activities targeting international and domestic audiences. For example, in February 2021, Twitter suspended 238 Iran-based accounts for alleged manipulative behavior during the first 2020 U.S. presidential debate (Reinl, 2018). Hossein Salami, the commander-in-chief of the Islamic Revolutionary Guard Corps (IRGC)1, also said that the corps has two thousand cyber battalions (Tabnak, 2022). These battalions work to fight the regime's enemies and dominate cyberspace for the regime's benefit. Cyber army activity peaked during the #MahsaAmini movement in September 2022 and the following months.

The #MahsaAmini movement was launched to protest against the brutality of Iran's morality police, who reportedly killed a young 22-year-old woman in custody. Police had arrested Mahsa Amini for violating the obligation to wear the Hijab. Mahsa Amini was reportedly subjected to severe physical abuse, resulting in brain damage and hospitalization. However, Iranian authorities denied these allegations. Amini passed away on September 16th, 2022, causing widespread shock and outrage among the public. #MahsaAminni emerged as a significant movement against the regime, with numerous individuals taking to the streets in various cities across Iran. This movement led to a record-breaking number of tweets, with over 500 million under the hashtag #MahsaAminni as of December 9th, 2022.

However, there needs to be more research on SMS campaigns in Iran. A growing, albeit small, number of studies focus on Iran's information operations (I.O.s) against foreign countries (Elswah & Alimardani, 2021; Howard et al., 2019). While research on false information and online incivility in Iran is rare, there are some works on bot activism in the Iranian Twittersphere (Farzam et al., 2022; Kargar & Rauchfleisch, 2019; Mazoochi et al., 2023). Nonetheless, most of these studies repeat what we already know about the negative impact of bots on social media communication. They do not contribute to a new understanding of how and by whom SMS happens in a non-democratic society.

In order to address these gaps, the following research questions (R.Q.s) guide this study:

> **RQ1:** Who were the main actors in suppressing the #MahsaAmini movement on Persian Twitter?
>
> **RQ2:** Which tactics were developed and employed by bad actors to suppress the #MahsaAmini movement on Persian Twitter?

[1] IRGC is a multi-service primary branch of the Iranian Armed Forces, which is known as the ideological custodian of Iran's 1979 revolution.

**RQ3:** What were the targets of employing the above-mentioned suppressive tactics?

# 4. Methods

## 4.1 Data collection

We have collected data from Twitter's Academic API from September 15th to November 15th, 2022, during the first two months of the movement. Mahsa Amini was killed on September 16th, so we started data collection a day earlier to ensure the collection of relevant tweets. The first two months of protests were the movement's peak when many Iranians participated on a large scale. The total number of tweets with #MahsaAmini in Farsi (#مهسا_امینی) during this time surpassed 202,123,658 tweets. We concentrated on popular tweets, i.e., tweets with more than 1k likes in a day. Our data collection resulted in 78,088 tweets sent by 10,308 unique users. Having collected data, we have coded both users and their tweets discursively. Our interpretative approach is discussed below.

## 4.2 Human-driven qualitative and discourse analyses

We followed Social Media Critical Discourse Studies (SM-CDS) (KhosraviNik, 2017) as a convenient discursive approach to interpreting Twitter data. We refrained from using traditional discourse analysis approaches like critical discourse analysis (CDA), as the fluid, changeable, and non-static nature, location, and dynamic of the discursive practices in social media need a unique and appropriate approach, which SM-CDS provides us with one. SM-CDS model includes a horizontal context substantiation, which deals with the intertextuality among textual practices, and a vertical context substantiation, which links both the micro-features of textual analysis and horizontal context to the sociopolitical context of users in society. SM-CDS enables us to discursively investigate users' metadata and tweets simultaneously to reach the deeper layers of meaning-making on Twitter. The following paragraphs discuss how this approach informs the coding process.

Five coders have coded users and tweets in three rounds: inductive and deductive. Saldaña's (2015) two-step approach drove the first inductive rounds to qualitative coding. Per Saldaña's instruction, coders started their work by making start lists (for users and tweets separately). A start list in provisional coding is generated based on literature reviews related to the study, the study's conceptual framework and research questions, previous research findings, and the researcher's previous knowledge and experiences (Saldaña, 2015, p. 144).

The start lists were then developed into provisional codesheets by closely and systematically reading the tweets. Codebooks were used as a baseline for the second inductive round. In this round, coders coded the whole samples based on the pattern coding (Saldaña, 2015, p. 209). In this step, coders were free to alter the provisional codebooks, for instance, when a new theme emerged, or an old one was not representative enough. The final codesheets were framed after all these steps.

In inductive steps, coders relied on the SM-CDS model to investigate accounts/tweets and their connections with sociopolitical discourses in Iran. SM-CDS enabled us to go beyond the text's manifest meaning to interpret social media's discursive layers of signification and intertextuality. In doing so, coders did not restrict their interpretation to the selected users and tweets. All meta-data were analyzed, including users' profile information and images, videos, and hyperlinks embedded in a tweet. Moreover, coders referred to Twitter to read replies and discussions beneath each tweet. Coders also took qualitative notes on findings and classified tweets into specific categories.

Having finished the inductive steps, coders used the final codesheet to quantify the codes across all sample users and tweets. This step was closed, and coders were prohibited from changing or altering the codebook. Of course, we did not solely rely on these numbers; we also used the qualitative notes and memos in the previous rounds to understand grounded practices better.

Coders followed the above approach to code users first. The initial aim of coding users was to find pro-regime accounts. As a result, 1,388 accounts were classified as such. The coding process continued with coding users based on their political orientation (pro-regime radical, pro-regime-moderate, and suspicious), ICB, type (e.g., journalist, elite account, etc.), and identity (e.g., real). We presented all variables and codes with their operationalized definitions in Appendix 1.

Then, we extracted all tweets from these users (N= 6,170). As discussed above, all these tweets have gone through a close and delicate process of discursive coding. Tweets were coded for the disruptive practices (e.g., incivility and its types) and the practice's target (e.g., discrediting activists). In the final stage, we calculated intercoder reliability using Krippendorff's alpha (Lombard et al., 2002). The lowest alpha was 0.73. Thus, we concluded that the codes are reliable and valid enough.

In the next section, we present the findings to answer RQs.

## 5. Findings

We start by investigating the characteristics of pro-regime users. Then, we will turn to the practices and mechanisms those users employed to suppress the movement.

### 5.1 Who were the bad actors?

Our analyses of pro-regime users (PR-Users) reveal that most of the accounts on this side belonged to radical users, real accounts, and elite users. These findings contradict most of our hypotheses and previous research (Forelle et al., 2015; Keller et al., 2020). Out of three PR-Users groups, the majority of accounts (%74,9) were pro-regime radical users. Suspicious accounts that are believed to be at work to infiltrate anti-regime groups consisted of merely 18,31 percent of accounts. Finally, pro-regime moderate users comprised a small share of accounts, %6,79. The dominance of radical users is also evident in the quota of shared tweets. They shared %86,6 of all tweets, while suspicious accounts followed them with %9,82, and moderate accounts only shared %3,58 of popular tweets. The majority of radical accounts show the tendency of Iran's regime to polarize space. This tactic would allow them to justify violent actions and suppress protesters in the streets. However, the low number of suspicious accounts is not anticipated. The next finding confirms this result.

Histogram 1 below shows the percentage of ICB probability among PR-Users.

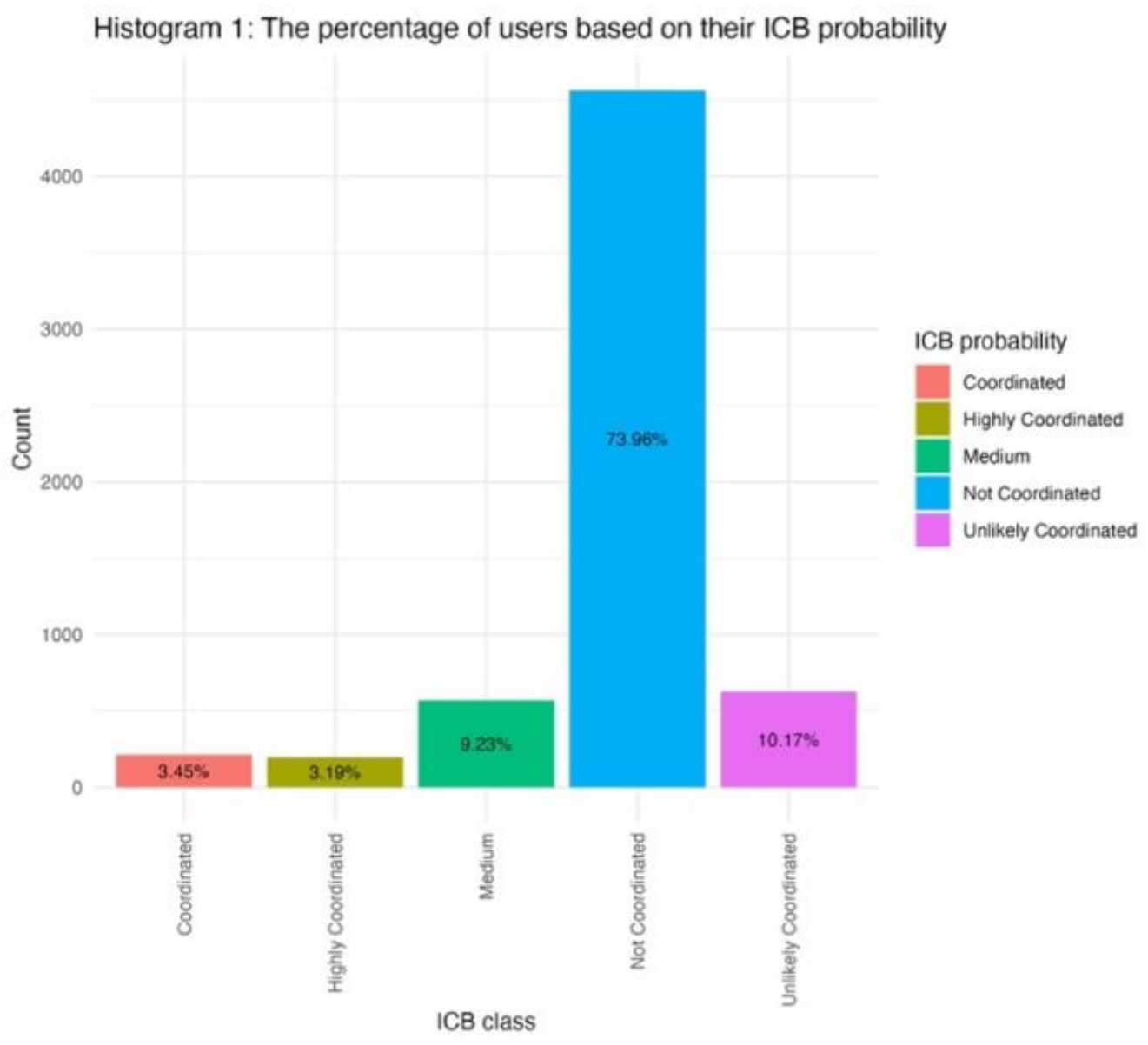

**Histogram 1**

Histogram 1 indicates that most accounts were not probably coordinated (in total: %84,13). Given that we went beyond the dominant approach in the existing literature in investigating bots as algorithmic accounts and considered other types of ISMA, this result shows that most of the popular users on the pro-regime side during the #MahsaAmini movement were not bots or other malicious types of accounts. This finding reveals that while existing research concentrates on bots in social media suppression, such a perspective is not necessarily true everywhere, particularly in Western societies. In an authoritarian regime like Iran, the regime did not employ many algorithmic agents. Instead, individuals who strongly support the regime are propelling its propaganda. It could be contingent on several facts. First, the contingency and full control over society may make authoritarian regimes needless of employing covert tactics and agents. They see no need to hide their suppressive activities. In addition, Neyazi (2020) indicated that authoritarian regimes can force government employees to work as bad agents by promising them promotions and other incentives. In this sense, employing humans may cost the regime even less than orchestrating bots' armies and make the bad campaigns seem more authentic.

We also analyzed the extent to which PR-Users identities were clear and identifiable. Results show that most PR-Users used real identities, while a smaller portion were anonymous accounts.

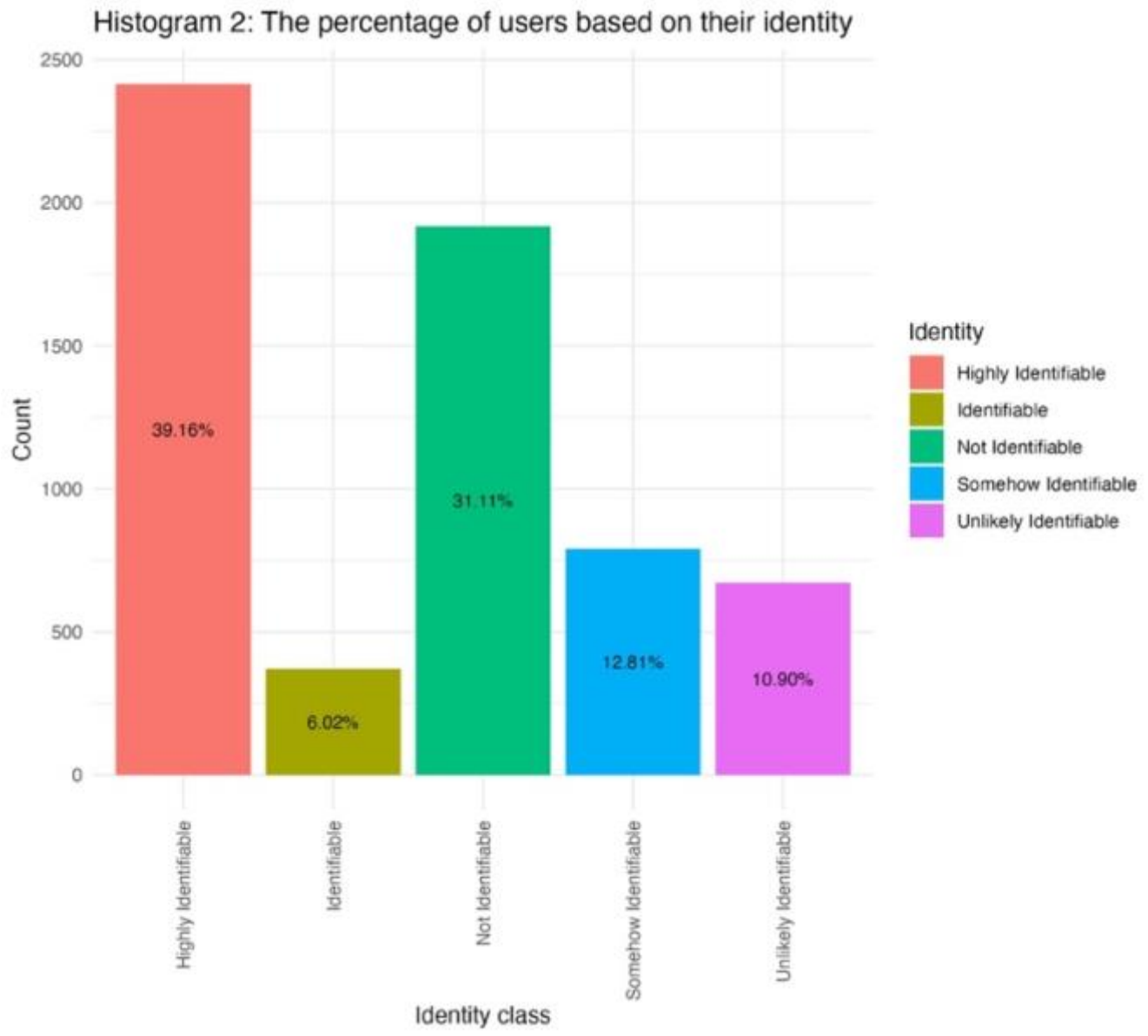

**Histogram 2**

The above histogram reveals that the identity of the majority of users was identifiable (in total: %45.18). On the other side, the percentage of accounts with non-identifiable identities was 42.1% in total. This finding confirms the above results, which show that there were not many automated accounts at work to suppress the #MahsaAmini movement.

We went even a step further to analyze the identity types of PR-Users. Histogram 3 below indicates the percentage of each identity type.

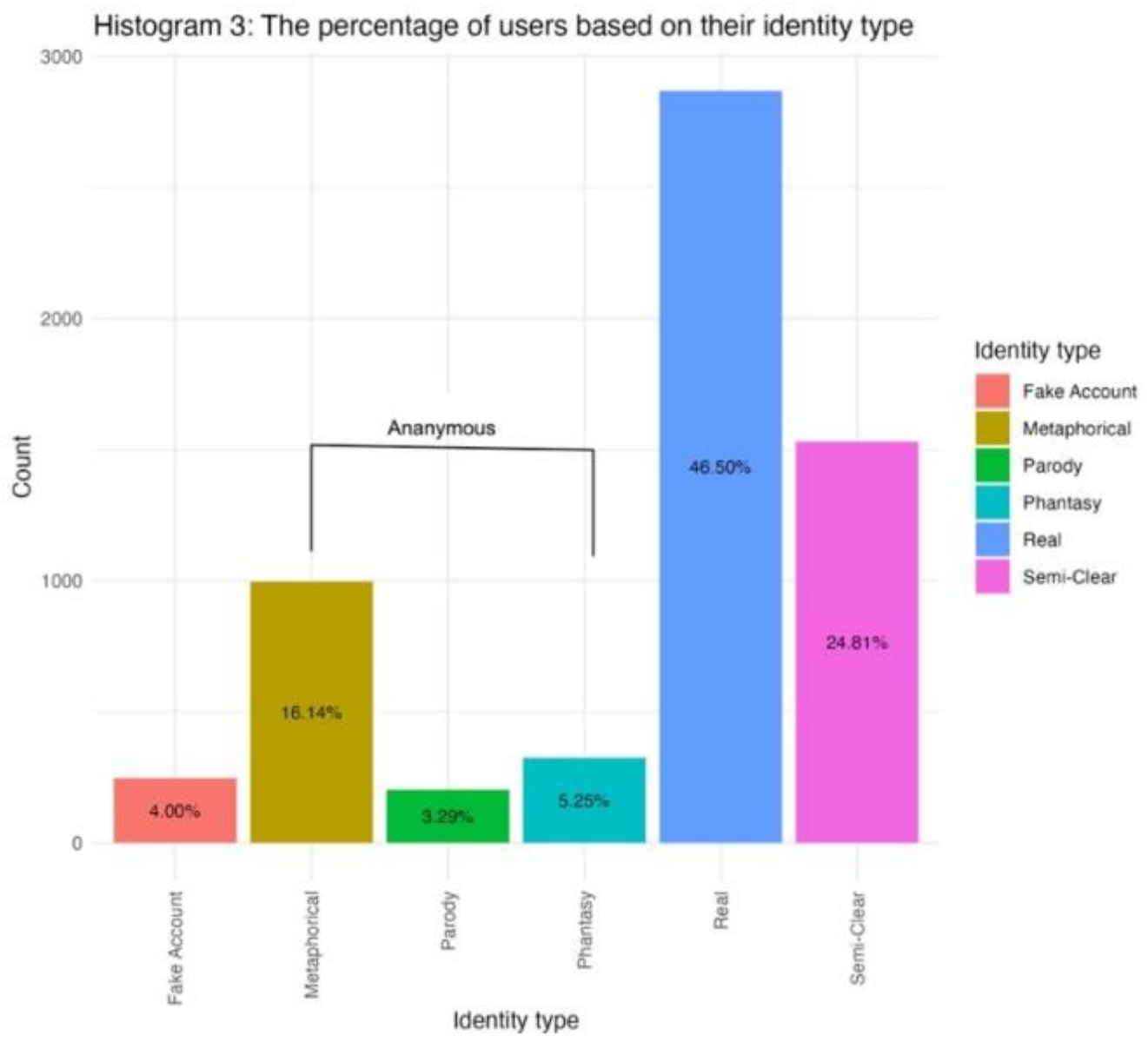

**Histogram 3**

This histogram shows that %46.5 of PR-Users used their real identities, and the identities of %24.81 are also semi-clear; they are not anonymous. Among anonymous identities, metaphorical usernames were more common. PR-Users mainly used them to refer to religious and revolutionary phenomena. For instance, there are multiple accounts with the prefix: *Haj*. Haj is a title used for religious figures, and in public, it refers to people who went to Mecca for a most famous Muslim ritual. This prefix is also a revolutionary one that pro-regime figures use to call each other. The diagram also reveals that there were only %4 fake accounts, which shows they were unpopular.

The final variable on the users' side was user type. Results reveal that most accounts belonged to ordinary users, but elite users shared the most popular tweets. Histogram 4 below depicts the share of each user type and their tweets in the research sample.

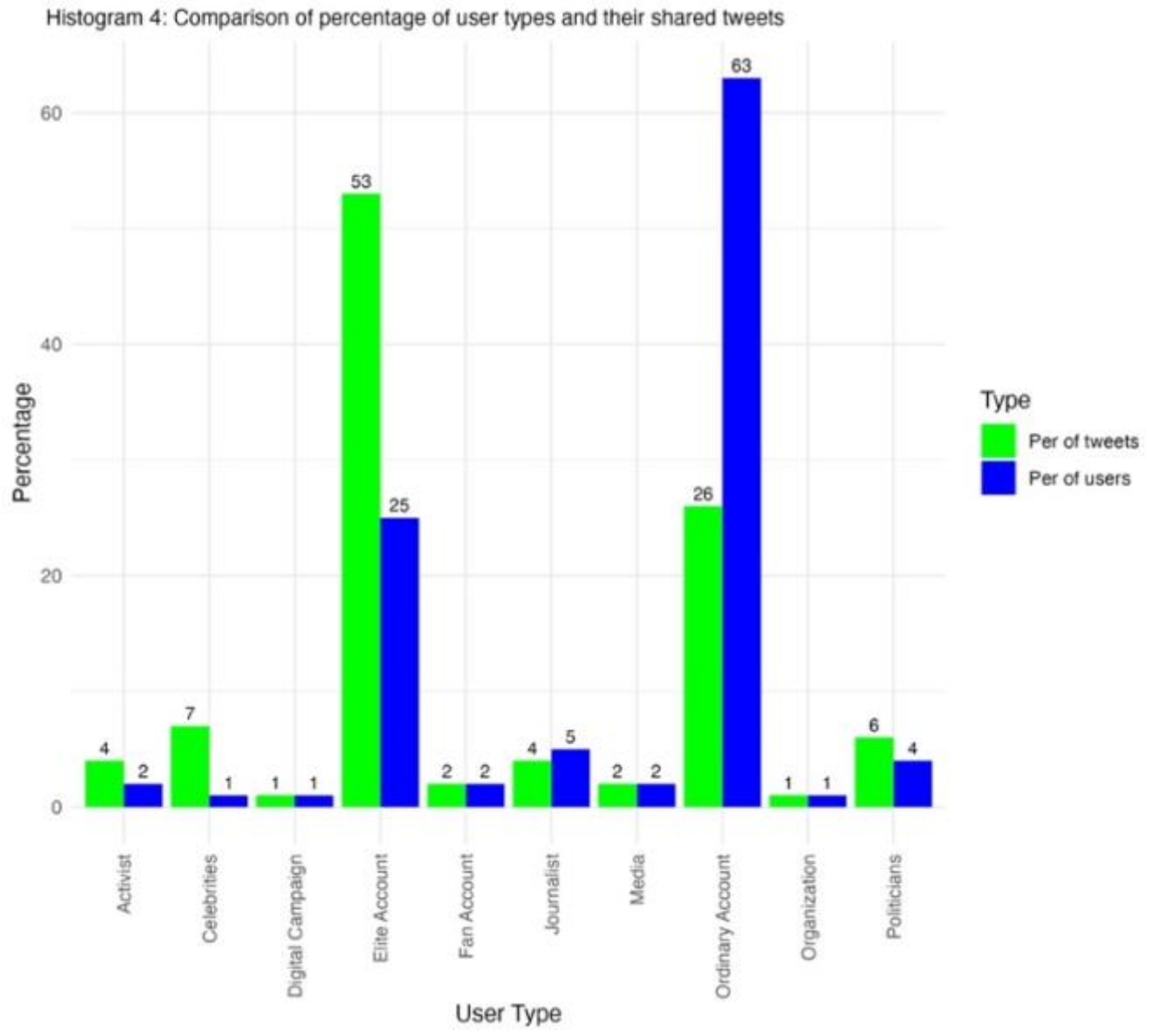

**Histogram 4**

Histogram 4 shows that while %63 of users were ordinary accounts, they only shared %26 of popular tweets. On the other hand, elite accounts consisted of 25% of users who shared %53 of tweets. Other types failed to rise to prominence. In particular, politicians', journalists', and the media's shares of users and popular tweets were insignificant. The results confirm that the Iranian regime relies more on individual agents than official figures like journalists and media outlets to suppress the movement. Of course, there was an interplay between media, individuals, and other user types in suppressing the #MahsaAmini protest, but elite accounts played the leading role. In this way, the regime could exonerate itself from killing the movement officially. In addition, since the regime blocks Twitter, the lower presence of official accounts could be reasonable. Finally, the reverse relation between ordinary/elite accounts' share of users and tweets could be contingent on the fact that many ordinary accounts are on Twitter only to amplify and propagate the elite accounts' propaganda. That could justify ordinary accounts' lower number of popular tweets while their users surpassed elite accounts.

This finding leads us to a significant point. As we will discuss later, our research shows that PR-Users' activity was most likely organized to a high extent but not necessarily automated. It challenges the dominant approach in the existing literature, which focuses on algorithmic intervention in social media activism in Western countries. The following section provides a novel understanding of social media suppression beyond computational propaganda.

## 5.2 The practices of suppression: Incivility, false information, and other tactics

In this section, we answer RQ 2. Our discursive analyses yield 12 suppressive tactics used by PR-Users. Histogram 5 below illustrates the top ones with at least %5 of the total sample (see their complete list in Appendix 1; for the statistics of all tactics, see Appendix 2).

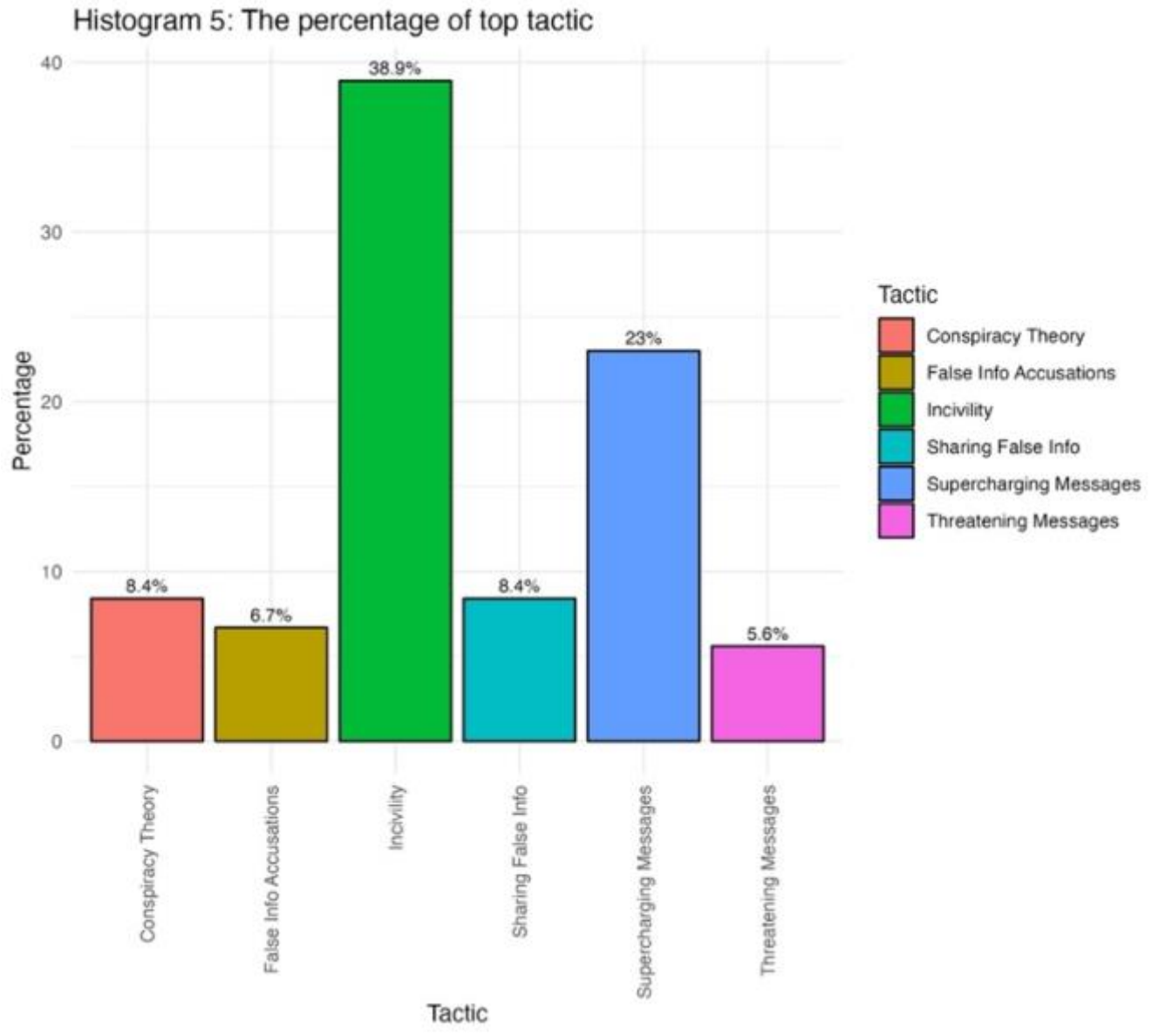

**Histogram 5**

The above histogram reveals that the main suppressive tactic was incivility, with almost %40. 'Supercharging messages' follows it with %23. Next, 'conspiracy theories' and 'sharing false information' are among the top suppressive tactics, with %8.4 for each. These findings indicate that the heavy emphasis on false information as a primary tactic of suppressing social media activism is not necessarily valid. Incivility, by far, was the major tactic in suppressing the #MahsaAmini movement. While the existing literature mainly ignores it as a suppressive tactic, our analyses show that PR-Users employ uncivil speech in their tweets to silence dissident users and turn the page in their favor. They employ uncivil speech to keep the regime's hegemonic discourse intact and prohibit dissident ideas from weakening it. It can be said that online incivility is the negative side of preserving the regime's discourse. For instance, PR-Users substituted the word 'life' in the movement's slogan: Woman, life, liberty with 'debauchery,' which is pronounced similarly to life in Farsi (in Farsi: Life is زندگی and debauchery means

هرزگی). In this way, PR-Users aimed to minimize all dissident citizens' wills to sexual relationships and then stop people from supporting the protesters.

PR-Users use another common uncivil tactic. They used riots (in Farsi: اغتشاشات) instead of protests (اعتراضات) in almost all of their tweets. In this way, they tried delegitimizing the protests by associating negative meanings with them. Substituting these words was a rhetorical tactic to depict protesters as non-rational hooligans and wild people, justifying the regime's violent response. We will discuss the aim of using profanity language in pro-regime tweets later.

We went further to dig deeper into the different types of incivility. The following histogram illustrates the percentage of each class of incivility.

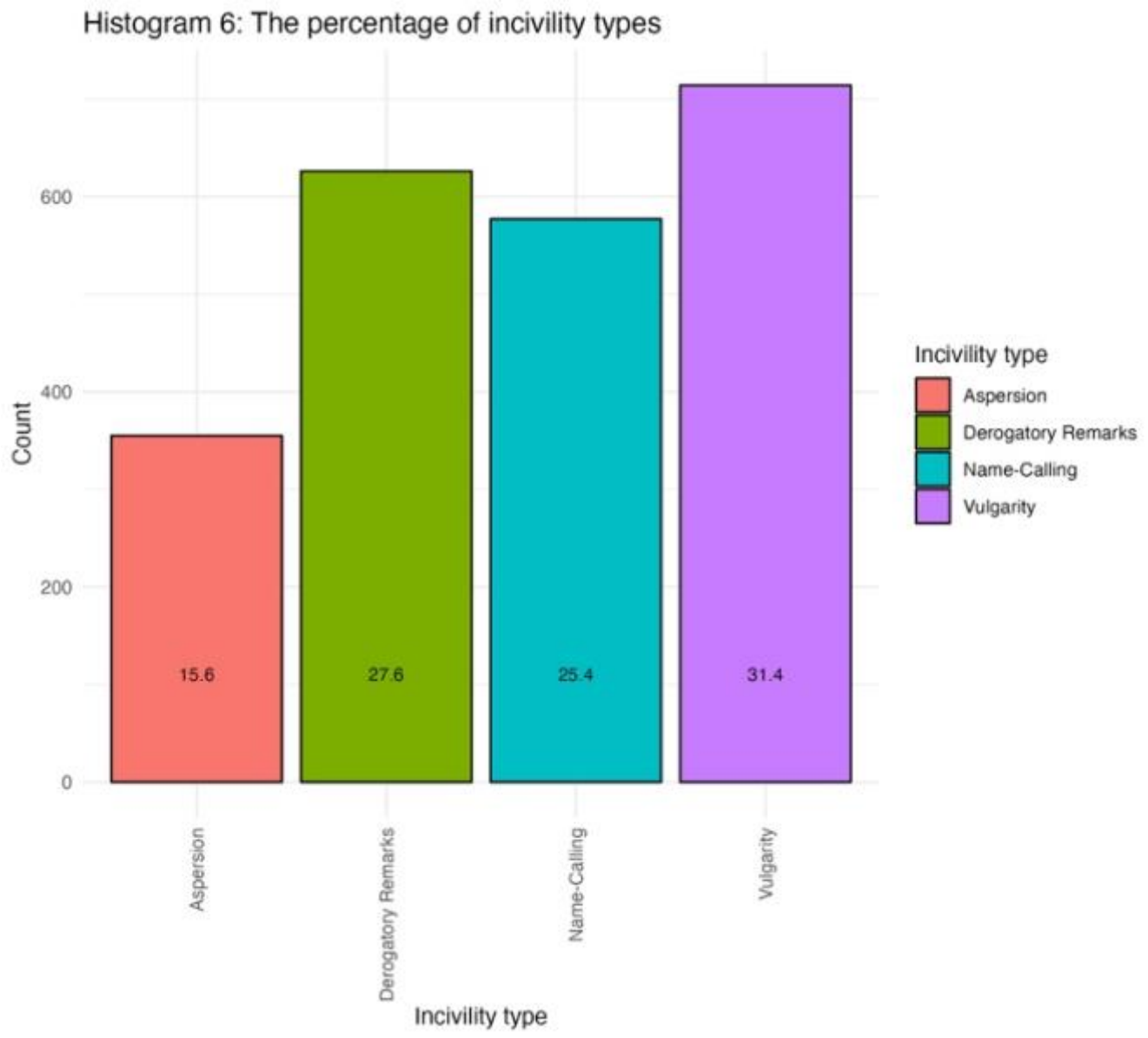

**Histogram 6**

Histogram 6 shows that aspersion was the less popular, while there was no significant gap between other incivility types in uncivil messages. It was expected that vulgarity would rise to popularity as an incivility tactic since using an insulting word solely could make a tweet vulgar. A significant recurrent theme in vulgar tweets was the same as the tactic mentioned above. PR-Users repeatedly used words like 'prostitute,' 'whore,' 'nudes,' and so on to discredit the protesters saying that they only disagreed with the regime because they wanted a free sexual life. In order to make this tactic more effective, PR-Users developed a rhetoric theme. They compared the 1978 revolution, referring to it as 'the revolution of nobles and protectors of the homeland,' with the #MahsaAmini movement, referring to it as 'the chaos of prostitutes and advocates of depravity.' Since femineity and freedom of women were at the center of the movement's aims, PR-Users tried to deny them without showing their true intention. They used 'prostitutes' as a metaphor to denote protesters; it made discrediting them and their wills easier. In addition, it was a tactic to support the anti-feminist nature of the regime's discourse in a covert way and marginalize other controversial issues like economic demands, freedom of political activities, and speech. In this way, vulgarity was an act of suppression.

The aim of pejorative speech messages was mainly to discredit activists but in a different way. PR-Users mainly embedded pejorative words in their tweets to criticize the protesters' intolerance and paradoxes. In this way, PR-Users wanted to challenge the anti-regime users' incapability and the emptiness of their claims. For instance, @sabeti_twt, a radical pro-regime figure, used the fact that Mohsen Namjo and Mahnaz Afshar, two famous Iranian artists, were kicked out of anti-regime protests abroad to say it shows that dissidents are intolerant. So, they will be totalitarian once they take power.

The aim of name-calling messages is most evident. These tweets directly attacked known and famous anti-regime individuals as well as other groups of protesters to discredit them.

Finally, aspersion was a tactic to question the whole idea of the movement and the activities of protesters. A rhetoric practice here was the appeal to pity fallacy. In numerous tweets, PR-Users argued that protesters fired and toppled the flag of Iran to depict the regime and themselves as victims. In another rhetoric tactic, PR-Users used ISIS (Daesh) as a metaphor to argue that protesters' behaviors resemble ISIS's violent activities. 'Homeland Daesh' was a metaphor that repeatedly appeared in PR-Users tweets. Using this metaphor, PR-Users ascribed negative meanings to anti-regime people and depicted themselves as peaceful and rational. This depiction is significant since the regime has been criticized for its violent response to citizens' democratic demands in recent years.

PR-Users used examples of protesters' violent behavior to support their accusations. Of course, protesters did violent activities occasionally, but the scale and institutionality of their violent behavior in comparison with the regime's brutal response were questionable. Ignoring these facts, PR-Users used aspersion as an atactic to intensify anti-regime people's violent behavior and mitigate the scale of the regime's violent activities.

While we understand incivility as the negative side of protecting the regime's discourse, 'supercharging messages (SMs)' are its positive side. The main aim of incivility was to prevent dissident citizens from weakening the regime's hegemonic discourse. On the other hand, SMs aim to connect the regime's ideology with desirable meanings. PR-Users shared many supercharging messages to refresh their spirits, propagate the regime's power, and blur the scene to conceal the voices of dissident citizens.

We found three main categories of SMs. First is *astroturfing* messages. This code refers to identical tweets that follow a similar pattern, shared by different users at different times. For instance, many PR-Users posted messages asking their fellow users to recite *Salawat*. Salawat is an Islamic ritual to send blessings to the prophet and show Allah (God) their respect for Muhammad. However, it has discursive meanings beyond its pure religious one. First, there is a remarkable difference between the phrases Shiites and Sunnis use for sending Salawat. While Sunnis only send it to the prophet, Shiites send it to him and his family. His family includes Ali, who Shiites believe should have been the immediate heir of Muhammad, but Abu-Bakr, the first Khalifa of Sunnis, stole it from him. This is arguably the most significant separating point between Shiites and Sunnis. All these battles and disagreements are hidden in Salawat. In this sense, Salawt for Shiites is a way of distinction and underpinning the Shiites' discourse. More than that, the Salawat took another meaning in the Islamic Republic. Many regime affiliates and believers use it to show they have the majority and raise their spirits. On some occasions, like politicians' speeches, when they want to silence dissident voices, they ask the crowd to recite Salawat. Thus, Salawat is a way to send a message to each other not to be frightened because we have the majority, and we will be victorious. PR-Users populated Twitter with such messages during the movement.

The interesting point is that while such messages were thematically identical, they were not exactly the same. It shows the probability of the fact that pro-regime activities were organized. However, those messages were sent by real identities, mainly ordinary accounts. There was a similar theme to Salawat's messages. PR-Users shared tweets admiring Iran's leader, Khamenei, as the symbol of the regime, with some poems or other literary messages. These messages were formally different to a large extent, but their themes were similar. We understand such messages as 'idea astroturfing,' which differs from 'whole message astroturfing.' The existing literature to date focuses on astroturfing messages as similar posts with minimal differences. However, it is a tactic for bad actors to share formally different messages but thematically the same. It makes identifying the cluster of bad actors by automated methods more challenging. As we discussed above, the Iran regime only employed a few ISMA during the movement. While such accounts could be tasked with sharing similar messages, principals of the suppressive campaign during the #MahsaAmini movement asked actors to share tweets with pre-defined content but in their own style and forms.

The next theme in SMs messages was bombastic tweets about Iran's power, in particular its military capability and advancements. This way, PR-Users tried to redirect the debate from domestic issues to international challenges and, at the same time, threatened foreign countries, mainly Israel and other Persian Gulf countries. Such messages were also a vehicle to boost their mood. The hostility against 'the West' is an inseparable part of Iran's ideology, and PR-Users reproduced this theme during the movement to not only suppress the movement but also entrench the regime's ideology. A clear example of such messages is the tweets regarding the *Hypersonic missile*, a weapon capable of travelling at hypersonic speed, which Iran announced that it built in November 2022.

The final theme in this category was normal tweets about everyday life. While such tweets seem not to be suppressive, they were indeed. They aimed to depict everything as normal at the time of the biggest crisis in the Iranian regime's history. These tweets mainly endeavored to limit the impact of the movement and accuse dissident people of being liars. At the same time, these tweets were a way to increase pro-regime citizens' confidence and trust in the regime. The main targets of these tweets were normalization and depoliticization. While the gap between the following tactics, false information, and conspiracy theory (CT), with the above-discussed tactics, is significant, they were still at PR-Users' disposal. The prevalence of CT messages and the high popularity of SMs tweets show that traditional propaganda tactics are still crucial for bad actors. In an authoritarian regime like Iran, the regime did not solely resort to sharing false information on an automated and large scale. Our research shows that old-school approaches, such as bombastic messages and sharing conspiracy ideas, proved to be as desirable as sharing false information. In line with the anti-Western hegemonic discourse of the Iranian regime, conspiracy tweets argued that the movement is a ploy orchestrated by Western intelligence agencies. It was a way to discredit the movement, employing pre-established discursive themes. The other example of CT messages was found in tweets following the *Shahcheragh* massacre. On October 26th, 2022, a terrorist attack happened in Shahcheragh, a Shiite pilgrim in Shiraz, Iran, which ISIS took responsibility for. However, many protesters accused the regime of being behind the attack. On the other side, PR-Users took the incident to blame protesters, arguing that the unstable situation in the country paved the way for terrorists. They repeatedly argued that it is the protesters whose blood of victims are on their hands.

Our findings also show that sharing FI and FI accusations was another means employed to suppress the movement to some extent. However, they were less widespread than the tactics mentioned above. Here, we should remember that we separate CT from FI regarding the research aims. If we consider CT as a form of FI, as suggested by several studies (Kapantai et al., 2021), then the percentage of FI would increase. Nevertheless, our study reveals that this tactic was less prevailing than expected.

The last dominant strategy of PR-Users was threatening messages. This tactic has three facets. The first category was tweets that directly threatened protesters of being killed, tortured, or kidnapped. PR-Users employed a metaphor here sarcastically to mock protesters at the same time threatening them. *Gunny sack (in Farsi: گونی).* When a protestor was arrested, PR-Users employed this metaphor to say s/he had been put in the gunny. As discussed earlier, the next theme was threatening foreign countries as part of supercharging messages. PR-Users' final rhetoric practice was again the Daesh (ISIS) metaphor. They argued that if the regime were toppled, ISIS would occupy the country. They shared many tweets with images of ISIS forces in the streets of occupied countries like Syria, disseminating the fear of what would be Iran's future without the Islamic regime. The below image is an instance of this class sent by @haj_haydar, a radical elite user.

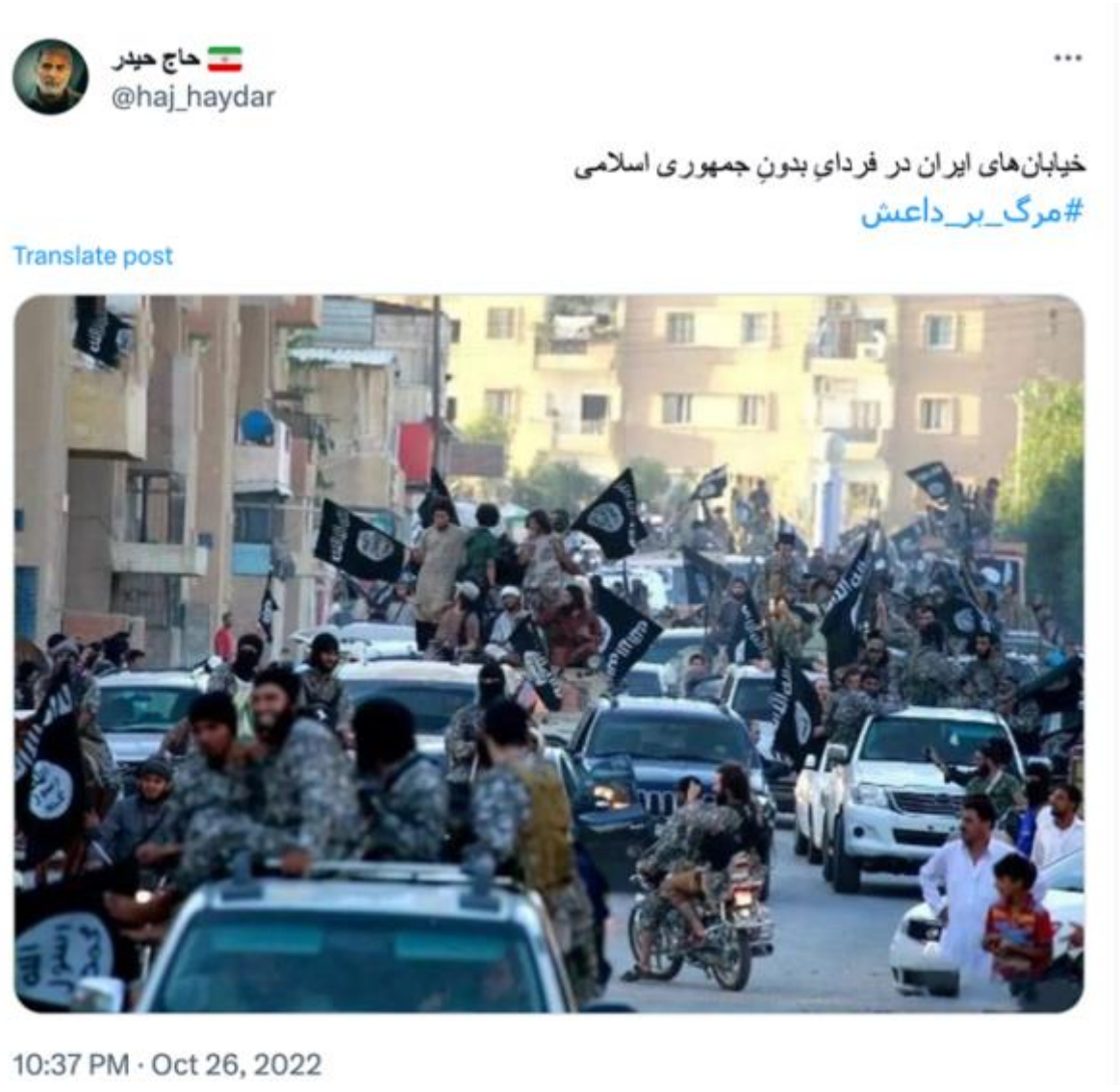

**Image 1:** a tweet by a radical elite user

### 5.3 The co-occurrence of tactics

In this section, we explore the co-occurrence of suppressive tactics. Findings reveal 46 tactics co-occurrence, with a total frequency of 1,159. However, most of the co-occurrences have a small value. Table 1 depicts co-occurrences with more than %10 of all instances.

**Table 1:** Top co-occurrences of tactics

| Tactic 1 | Tactic 2 | Frequency | Percentage |
|---|---|---|---|
| Incivility | Conspiracy theory | 219 | 19 |
| Incivility | Sharing false info | 188 | 16 |
| Incivility | False info accusations | 159 | 14 |
| Incivility | Threatening messages | 127 | 11 |

Table 1 shows that all top co-occurrences include incivility as a tactic, which was expected due to the high prevalence of this tactic. Furthermore, since only using a profanity word makes a text uncivil, it makes sense that incivility combines with other tactics to a more significant extent. In fact, incivility is a grounded tactic, and its combination with other tactics could intensify the suppression level of the text. Our analyses reveal that incivility mainly combines with negative tactics, particularly CT and FI As Table 1 indicates, the combination of incivility and SMs, as the positive side of suppressive tactics, is not among the top ones. Besides this, the frequency of top tactics in Histogram 5 could be used to predict the co-occurring tactics precisely.

Since incivility is mainly aimed at silencing dissident voices, its combination with other suppressive tactics is probably a way to prevent counter-argumentations; as a result, sharing CT and FI with less resistance. For instance, @Gandomizadi, another elite user, has posted the same photo as image 1 with a different caption, combining CT with incivility: *You will be choked up when you open the window one day to chant Woman, Life, Liberty and happen to see this scene*. This tweet tries to leave no room for objection by insulting all protesters. In addition, tweeting the same photo with a different caption confirms the probability that PR-Users activities are organized. Searching the same tweet shows other PR-Users tweeted the same image with the same caption several times. These PR users are mainly ordinary ones with fewer followers, and their tweets have received less attention. However, they tried to overwhelm Twitter with such messages. This trend has been seen in many cases, increasing the probability of an organized campaign. The pattern is an elite user who shared a Twitter that received many likes, and other ordinary accounts copied and pasted the same tweet. It is the form of Astroturfing that has been at the center of attention in the existing literature. We called it 'whole message astroturfing earlier.' (Clip 1 in supplementary material shows the Astroturf cluster of the above tweet).

## 5.4 The aim of suppressive tactics

In this section, we investigate the target of suppressive tactics to answer RQ 3. Our analyses resulted in 5 main targets listed in Table 2.

**Table 2:** The frequency and percentage of targets

| Target | Frequency | Percentage |
|---|---|---|
| Attacking weak links | 96 | 1,9 |
| Discrediting activists (DA) | 1,337 | 25,9 |
| Normalization/depoliticizing | 622 | 12 |
| Redirecting debate (RD) | 2,977 | 57,6 |
| Stirring conflict within the anti-regime forces | 135 | 2,6 |

The above table shows that the targets of PR-Users suppressive tactics mainly were discrediting activists and redirecting debates. Findings confirm that PR-Users heavily tried to redirect the preliminary debate from the brutality of the regime and protesters' demands to interventions by Western governments, the violent behavior and paradoxes of protesters, the worst situation in Western countries, and the innocence of the regime. PR-Users also tried to attack the opposition leaders and other activists during the movement to limit the impact of the protest. We examine the distribution of targets across (all) tactics in Heatmap 1.

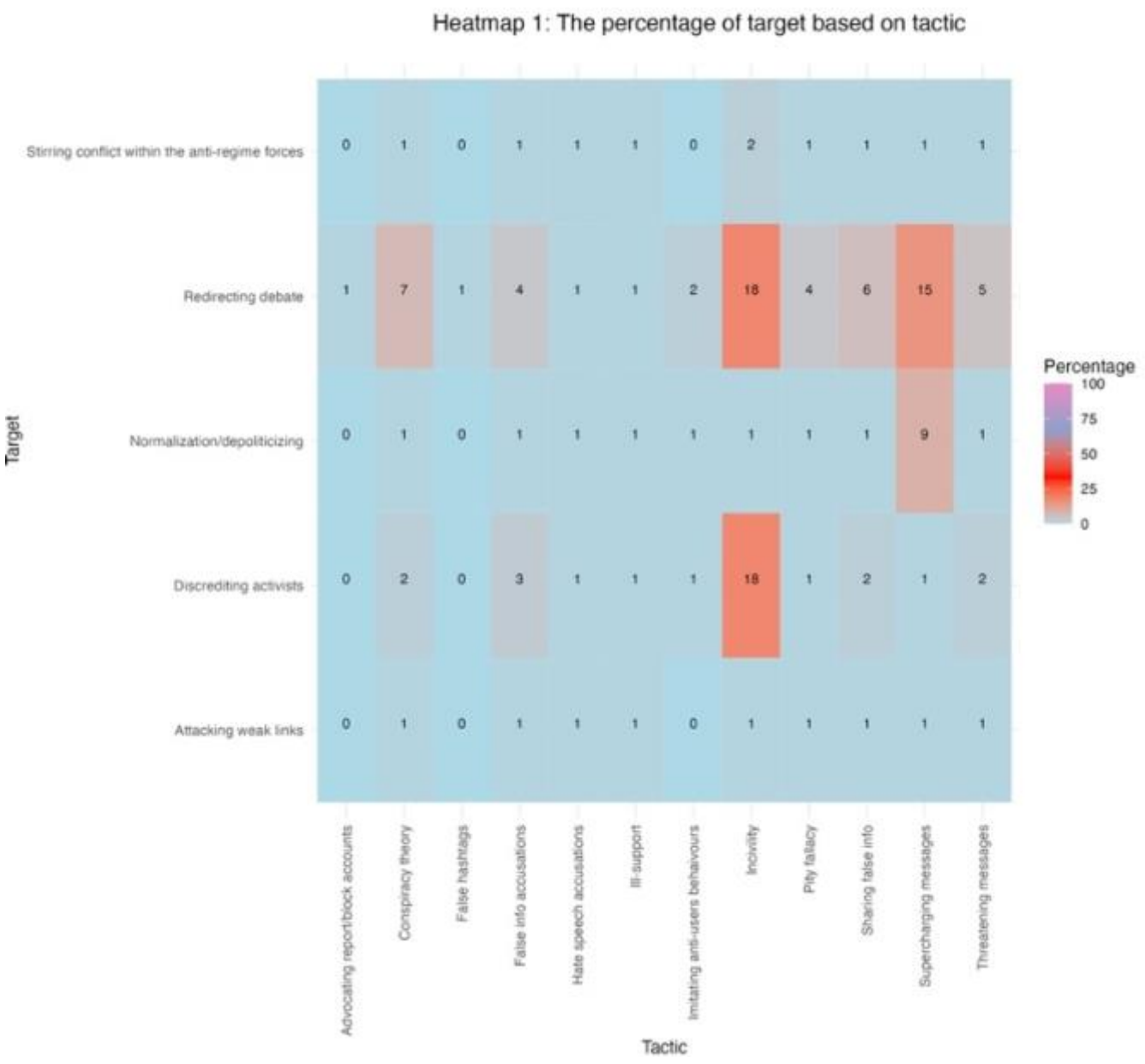

**Heatmap 1**

Heatmap 1 denotes that incivility was the main tactic in RD and DA. PR-Users mainly tried to redirect the debate by asking rhetorical questions. Using this tactic, they aimed to force anti-regime users to respond so they could not present their wills. They employed profanity language, aiming to boost the effectiveness of their messages. For instance, @famil2r, a pro-regime activist, shared an image of victims of a terrorist attack in Kabul, Afghanistan, and accused protesters of being silent about them. The tweet (see Appendix 3) combines dirty words and phrases and tries to redirect the debate to another instance. Another tactic in RD was SMs. Sharing positive messages about the regime and the leader was another way to redirect the debate. SMs also was a way of normalization. By this tactic, PR-Users try to depict everything as usual and deny any crisis or emergency situation. However, it seems that PR-Users knew that normalizing the crisis was not easy, so they put their energy mainly into redirecting the debate, which sounded more plausible at that time.

CT and FI accusations were other tactics in RD These strategies both aimed to provoke anti-regime users and force them to react and engage with PR-Users. In particular, as we see in Table 1, PR-Users mixed these tactics with incivility to make them more and more proactive. The eventual target could be wasting the dissident citizen's energy discussing with the regime's supporters, so anti-regime people would have less time and energy to protest against the regime.

PR-Users also used uncivil speech to attack activists to a large extent. However, their lessened use of FI in this regard contradicts the existing literature and our expectations. A reason for this could be the fact that PR-Uusers understood that the inaccuracy of FI could be revealed quickly and was less effective. In addition, they followed the same logic as we discussed above to make anti-regime people angry and make them interact with them, to waste their time and energy. A group of interest here was university students. Students have been a significant force in the history of anti-regime movements. During the #MahsaAmini movement, they were a big part of the protest. The regime attacked many universities to suppress them on the ground; PR-Users also tried to attack and mock them online. On October 2nd, 2022, the regime sieged and attacked the University of Sharif, a respected Iranian University, and arrested several protesters. At the same time, PR-Users shared many tweets sneering at the students. They used words like 'dump,' 'stupid,' and so on to call Sharif students, who are known as the elite people of the society. For instance, @elyas_ghanbari1, a pro-regime journalist, tweeted: *The holy recycle bins of the regime are worth more than vituperative students.* Of course, this tactic was also used to attack the movement leaders. For example, @sabeti_twt, a pro-regime TV host, attacked Golshifteh Farahani, an Iranian actress who lives abroad, by saying: *'Your sole skill is nudity.'* Insulting female activists was a common strategy to denounce them. In another instance, a pro-regime user attacked Nazanin Bonyadi, another Iranian actress, by sharing a photo of her and writing: *She is Nazanin Bonyadi, the new leader of the revolution of prostitutes.*

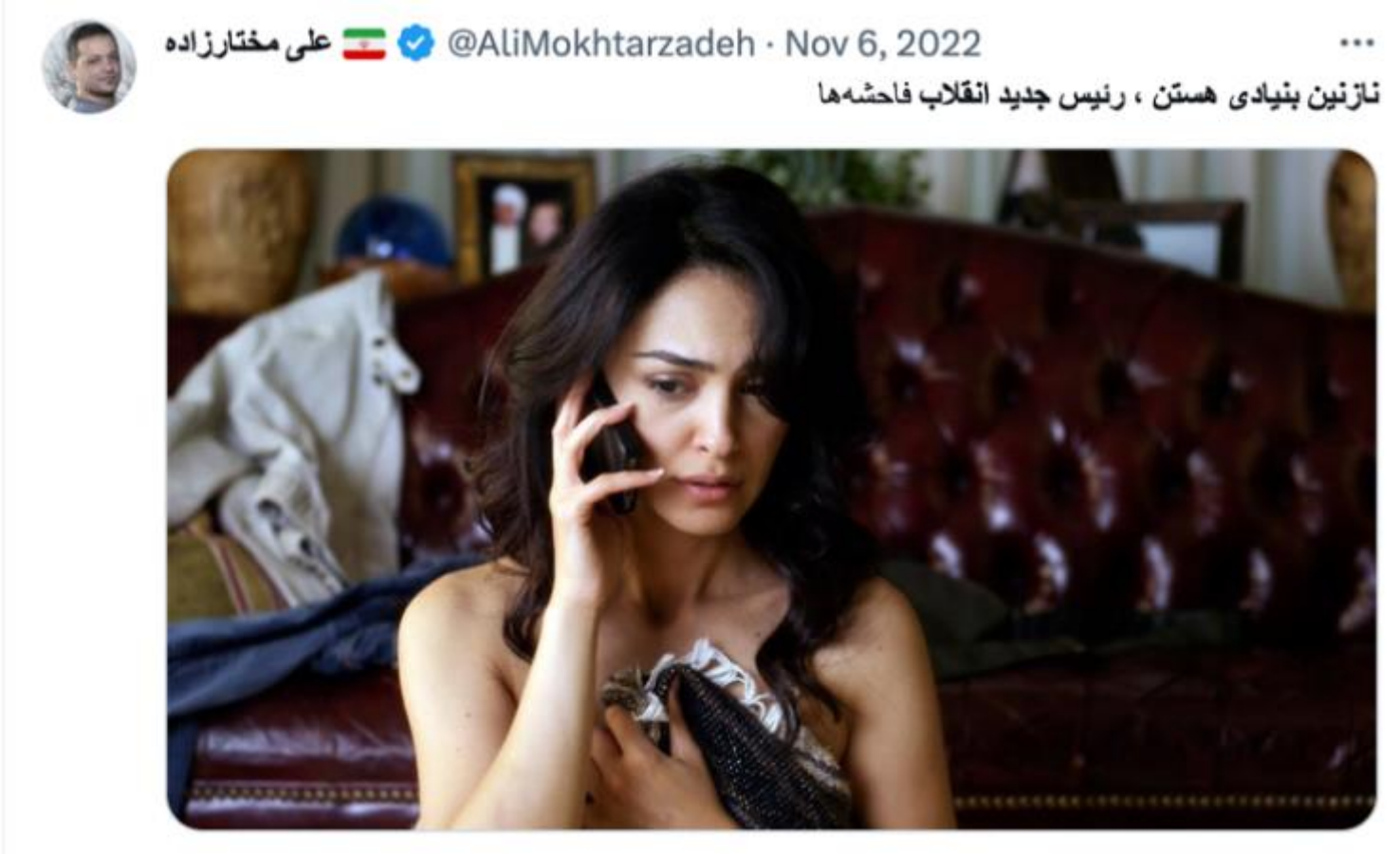

**Image 2:** A tweet to attack Nazanin Bonyadi

Findings show that the primary aim of FI was also to redirect debate. As an example, @ khorschid_t, a radical pro-regime fake account, shared a tweet that argued: *An hour before yesterday's riots, we talked and laughed with special forces. You don't dare to do the same with the Police in Germany. If these* *special forces were not here, riots would never let us walk freely in the streets by harassing girls!* This account uses an image of a blonde girl whose Hijab is certainly not acceptable by the regime. She argued that she is living in Germany. Through using such accounts, the regime tries to show that even Iranians who are living abroad and do not comply with the Islamic rules support the regime. In this tweet, the user has tried to redirect the debate by comparing the behavior of police in Iran and Germany. She also accused protesters of being harassers and labeled the protests as riots. Thus, she mixed incivility with different types of FI to redirect the debate from the protesters' demands.

In sum, PR-Users primary endeavor was to redirect the debate during the movement using incivility. By this, they tried to bring other and margin debates to the center of attention to minimize the scale and breadth of anti-regime users' demands. In many cases, they tried to change the victim and perpetrator positions to make dissident citizens furious, forcing them to interact with PR-Users and waste their time and energy.

### 5.5 Mapping users to tactics

In the beginning, we expected more diversity of users. Thus, we planned to analyze the relations between tactics and users. However, as discussed above, the research sample was dominated by elite/ordinary, non-coordinated, and real accounts. Nevertheless, we examined if there were any unusual relationships in the sample. Results indicate that elite/ordinary and non-coordinated accounts were responsible for prevalent tactics and targets. In all cases, elite accounts were prominent in spreading uncivil and supercharging messages. This finding is valid for other results. For instance, the role of fake accounts in repressing the movement or media/journalists in sharing FI was minor. We provide heatmaps confirming these observations in Appendix 2. Since there are no significant relations here, we do not discuss it in more depth.

## 6. Conclusion

This research enhances our understanding of the actors and practices of suppressing the #MahsaAmini movement on Persian Twitter. Given the scale and the movement's received international attention in 2022, it is necessary to study how the movement unfolded and repressed. In this sense, the current study extends the literature into social media suppression beyond Western countries, but this is not the sole contribution of this article. We provide fresh insights into social media suppression by focusing on an understudied and authoritarian regime: Iran.

The main takeaway of this research is to provide empirical evidence that social media suppression is not limited to FI and social bots. Thus, the limited scope of current existing literature, mainly conducted in Western democracies, gives us a deprecated image of SMS. This article argues for a more comprehensive understanding of suppression agents and practices.

First, our results reveal that the presence and role of automated agents in repressing #MahsaAmini were negligible. While the existing literature mainly focuses on social bots as the main force of online manipulation (Bessi & Ferrara, 2016; Howard & Kollanyi, 2016), we did not find their role significant. Our investigations also go beyond social bots to explore the role of other ISMA; the accounts are not necessarily algorithmic but coordinated like trolls and cyborgs. However, we found that the majority of pro-regime accounts were not coordinated. They were mainly real individuals with identifiable identities. This finding shows that the high focus on social bots investigating social media suppression agents could be misleading. The role of most social bots is indeed negative in multiple cases, but we need to go beyond them to investigate other ISMA. More than that, the real accounts could be sources of suppression. This study denotes that elite and ordinary users with identifiable identities mainly tried to suppress the #MahsaAmini movement.

The minor presence of bots on Persian Twitter contradicts other studies. In Russia, Stukal et al. (2017) discovered that the daily proportion of bots that were actively tweeting among Russian accounts was as high as 85%. Jones (2019) observed that during the conflict between Qatar and Saudi Arabia, at least 71% of the active accounts were found to be bots on some hashtags. In addition, the ISMA activity on Persian Twitter countered the high levels of bot activity discussed by scholars like Howard and Kollany (2016) during the BREXIT or Himelein-Wachowiak et al. (2021) during the pandemic.

There are some reasons for this finding. First, in most Western democracies, where the current studies heavily concentrate, it is not relatively easy for real individuals to use uncivil speech or share FI. Of course, it does not mean that there is no evidence of such malicious activities in democracies, but the scale of it is minor in comparison with authoritarian regimes. Citizens in democratic countries could legally sue someone for being uncivil or sharing inaccurate information. It is, more importantly, true for politicians or journalists. On the contrary, there are not only any consequences for people's malicious activities in non-democratic societies, mainly when they are in favor of the regime, but the regime also rewards these activities. It appears that there are no consequences for bad actors who participate in pro-regime suppressive campaigns.

In addition, using real people could be even less complicated and expensive in a non-democratic society like Iran. The regime has many employees working in its vast bureaucratic system. The promotion and rewards in the system are contingent on obeying the regime's rules. Thus, it is not a big challenge for the regime to task many employees with participating in the suppressive campaign in an organized manner.

Another significant finding is about the suppressive tactics. Our research reveals that against what the leading strand of literature argues, FI was not the central suppressive tactic in wrong agent disposal. Incivility surpassed it significantly. The reason for the dominance of incivility could be the same as what we discussed above: the fewer consequences for uncivil speech. In addition, our analyses show that PR-User's chief target of uncivil speech was redirecting debate. They employed profanity language to agitate dissident citizens, forcing them into discussion and interaction with PR-Users. They did the same to insult protesters and activists. Their eventual aim was to waste the energy and time of protesters in discussing the matters defined and dictated by PR-Users. Incivility was also another way to polarize the space and prevent any dialogue. Thus, any effort to challenge the regime rationally would be castrated. Our investigation shows that scholars should not limit their studies to FI Incivility as an act of suppression could be even more central to bad actors.

Supercharging messages and CT followed incivility. The results confirm that suppression happens in negative ways and in positive tactics. PR-Users shared many tweets admiring regime symbols like Khamenei to overwhelm the space and redirect the central discussion to the capability and desirability of the regime. In SMs messages, we also identified a novel tactic of Astroturfing. The existing literature mainly considered astroturf messages as identical texts shared on a large scale. However, we introduced the idea of 'idea astroturfing,' which means disseminating different messages with the same central idea. Those messages take more work to be identified and analyzed by automated methods.

Furthermore, the prevalence of incivility, SMs, and CT show that more traditional propaganda tactics are still at work in non-democratic societies. Technological advancements indeed make sharing FI and using social bots easier, but it does not mean that authoritarian regimes abdicate other tactics. Therefore, it is necessary to consider broader potential tactics in suppressing social movements.

Despite the contributions of this study, we should acknowledge some limitations here. First, this research is limited only to the first two months of the movement. Persian Twitter, as well as the political landscape in Iran, is fast-changing, particularly during and after the #MahsaAmini movement. Therefore, studying the movement over a more extended period could be beneficial in understanding the changes in suppressive tactics over time. On a related point, this study is limited to the visible layer of suppression. By this, we mean the apparent and immediate networks and tactics of suppression employed by the regime. However, during the research, we noticed that suppression could, and does, happen in more complicated and covert forms. For instance, the regime could employ fake accounts pretending to be anti-regime and infiltrate dissident communities. Some users could change their political orientation during the movement for several reasons, like being threatened by the regime.

We have noticed a surge in anti-regime monarchist users during the months after the movement. Monarchists support Reza Pahlavi, the son of the last king of Iran. While there are people who genuinely support the idea of monarchy in Iran, it seems that the regime uses many tactics to widen the gap among protesters by sowing discord and disagreement between them. We should acknowledge that these dynamics are out of the scope of this research. However, if we are to have a complete picture of the suppression of the #MahsaAmini movement, we must also study them. While we did our best to identify such suspicious accounts and behavior, and our results show they were minor, it could be an interesting topic for further investigation. In particular, a combination of social network analysis, natural language processing, and discourse analysis could be employed to dig deeper into the hidden layers of suppression on Persian Twitter.

This research also set other directions for further studies. While we did not analyze the activity of fake or suspicious accounts due to their minor presence, a deep dive into their practices and networks could be illuminative. Future studies can also use the list of suppressive tactics and agents we developed to investigate SMS in other contexts. The ideas of ICB, ISMA, and 'idea astroturfing' that we developed in this research could help other scholars go beyond the limited but dominant understanding of suppression agents and practices on social media.